\documentclass[runningheads]{llncs}
\usepackage[T1]{fontenc}
\usepackage{graphicx}
\usepackage{wrapfig}
\usepackage{subcaption}
\usepackage{enumitem}
\usepackage{zibtitlepage}
\usepackage{color}
\begin{document}
\ZTPAuthor{
  \ZTPHasOrcid{Inci Yueksel-Erguen}{0000-0002-1270-262X},
  \ZTPHasOrcid{Ida Litzel}{0009-0000-0166-7809},
  \ZTPHasOrcid{Hanqiu Peng}{0009-0004-1300-2854}}
\ZTPTitle{Integrating Large Citation Datasets}
\ZTPInfo{Preprint}
\ZTPNumber{24-10}
\ZTPMonth{Oktober}
\ZTPYear{2024}
\zibtitlepage

\title{Integrating Large Citation Datasets}
%
%
\author{Inci Yueksel-Erguen\inst{1}\orcidID{0000-0002-1270-262X}, Ida Litzel\inst{1}\orcidID{0009-0000-0166-7809} and Hanqiu Peng\inst{2}\orcidID{0009-0004-1300-2854} }
\authorrunning{I. Yueksel-Erguen, I. Litzel, H. Peng}
%
\institute{Zuse Institute Berlin, Takustrasse 7, 14195 Berlin, Germany \email{yueksel-erguen@zib.de}, \email{litzel@zib.de} \and National University of Singapore, Block S17, 10 Lower Kent Ridge Road, 119076 Singapore \email{penghanqiu@u.nus.edu}}
\maketitle              
\begin{abstract}
This paper explores methods for building a comprehensive citation graph using big data techniques to evaluate scientific impact more accurately. Traditional citation metrics have limitations, and this work investigates merging large citation datasets to create a more accurate picture. Challenges of big data, like inconsistent data formats and lack of unique identifiers, are addressed through deduplication efforts, resulting in a streamlined and reliable merged dataset with over 119 million records and 1.4 billion citations. We demonstrate that merging large citation datasets builds a more accurate citation graph facilitating a more robust evaluation of scientific impact.

\end{abstract}
\renewcommand{\labelenumii}{\arabic{enumi}.\arabic{enumii}}

\section{Introduction}
Evaluating scientific impact necessitates precise measurement of individual articles' impact, which is commonly assessed through metrics reliant on citation counts \cite{Hirsch2005,Garfield2006}. However, these metrics are subject to limitations, notably susceptibility to manipulation within the scholarly community \cite{FireAndGuestrin2019}. Recently, there has been a shift towards utilizing knowledge distilled from citation graphs rather than relying solely on citation counts \cite{Chen2023}. This shift mandates access to a comprehensive citation graph for more reliable measurement. 

A citation graph is a directed graph. Its nodes are the scientific publications. The arcs between nodes denote the citations. The academic research databases, also named as the citation datasets, store the academic publications and their reference lists needed to construct a citation graph. Yet, none of these datasets alone contains all academic publications \cite{Visser2021}. Thus, to construct a comprehensive enough citation graph, data in these big datasets must be merged in a consistent way. Recent studies in the literature on merging citation datasets either report limitations to working with large datasets due to memory \cite{Nikolic2024}, or report their results over a limited time span \cite{Visser2021}.

In this study, we focus on methods for constructing a comprehensive citation graph. We present a method for efficiently merging large citation datasets utilizing parallel computing. This method bypasses the need for database management systems or complex big data frameworks, promoting \textbf{widely accessible research}. Again, it extracts data as needed leading to a \textbf{quick start-up} without lengthy setup processes common to specialized database systems. It is also agnostic to data format allowing for application to diverse datasets, regardless of structure or volume.  

In Section \ref{Sec:Method} we present our method and report our remedies to handle significant challenges presented by big data. These include quality issues stemmed from semi-structured data lacking universal unique identifiers (UID). We report our implementation in Section \ref{Sec:Results}. Through meticulous deduplication efforts, we streamlined the merged dataset to a single consistent dataset. Our work led to a citation graph containing more than 119 million nodes and 1.4 billion edges. In Section \ref{Sec:Results}, we also show that this graph portrays inter-article associations more accurately than graphs derived from single datasets.

\section{Merging Big Research Datasets}\label{Sec:Method}
Merging research datasets poses several key challenges. One primary objective is to create a comprehensive and accurate citation graph by combining the publications recorded in individual datasets. This involves finding the union of publications and consistently merging their reference lists.

Let $D_s$, $s=1,2$ be the datasets we would like to merge. $P_s=\{p_{s1},...,p_{sn_s}\}$ is the set of scientific articles in dataset $D_s$ where $n_s$ is the number of records. Attribute set of $D_s$ is $A_s=\{a_{s1},...,a_{sm_s}\}$ with $m_s$ being the corresponding number of attributes. Each tuple in $D_s$ is a publication record denoted by $\pi_{si}=({p_{si}}^{a_{s1}},{p_{si}}^{a_{s2}},...,{p_{si}}^{a_{sm_s}})$, where every entry is an element of $P_s \times A_s$. Our aim is to find $P=\{p_1,...,p_l\}$ such that $P=\big (\bigcup_{s=1}^2{P_s}\big )/ {\sim}$ holds under a carefully designed equivalence relation $\sim$ and compile the corresponding citation sets.

\begin{figure}
\includegraphics[trim={0cm 5cm 0cm 5cm},clip,width=1\textwidth]{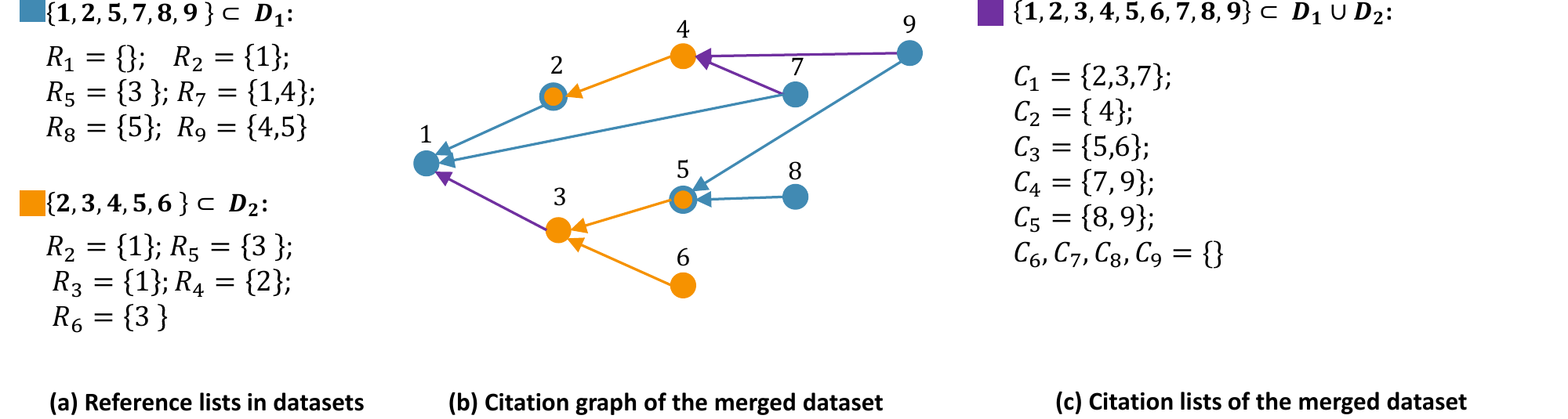}
\caption{Merging two datasets} \label{Figure_MD}
\end{figure}

As an illustrative example, consider Fig. \ref{Figure_MD} depicting two datasets, $D_1$ and $D_2$, containing articles $p_{si} \in P_s = \{1,...,9\}$ and their references $R_{p_{si}}$. Blue (yellow) nodes represent articles in $D_1$ ($D_2$), while bi-colored nodes signify articles present in both datasets. Blue (yellow) edges depict citations added based on information from $D_1$ ($D_2$). The crucial purple edges require matching references across datasets to the corresponding articles in the merged dataset.

As the records of existing academic research datasets do not completely match to another \cite{Visser2021}, merging datasets is necessary for a comprehensive citation graph. The completeness and consistency of such a citation graph hinges on detecting all matched records/references of the input datasets. However, achieving this level of accuracy necessitates performing several challenging tasks.

\textbf{Record matching} is the first challenge. To maintain uniqueness of records, it's essential to accurately identify records addressing the same article across input datasets. In Fig. \ref{Figure_MD}, this corresponds to correctly detecting and matching the records in $D_1$ and $D_2$ representing the articles $2$ and $5$.
The other challenge is \textbf{reference matching}. This involves identifying references in each dataset that correspond to articles within the merged dataset to construct the citation lists, denoted as $C_{p_{si}}$ in Fig. \ref{Figure_MD}, consistently. For example, in the figure, $3$'s attribute pointing to the reference to the article $1$ in $D_1$ must be detected correctly.

The lack of a universal UID for records complicates these tasks. Research datasets rely on publication metadata, including title, journal name, ISSN, authors, publication date, and reference list. While Digital Object Identifiers (DOIs) offer a potential solution since their adoption after their announcement in 1997, they are not ubiquitous across all articles or datasets. Thus, text-based data attribute comparison is needed for record and reference matching.

This highlights the challenge posed by non-indexed text fields in research datasets, often stored in semi-structured formats like XML or JSON. These fields are susceptible to data entry errors, further compromising data consistency, particularly in open data sources relying on publisher-provided information. While the vast size of open datasets is valuable, it necessitates the development of efficient methods for extracting relevant and reliable data attributes for effective analysis. Large research datasets demand efficient ways to select key data attributes for merging them into a consistent and complete citation network.


In this work, we prioritize the development of a methodology utilizing widely available software tools and open-source libraries. Again, we ensure compatibility across various dataset formats. Hence, we focused on techniques avoiding reliance on proprietary database management systems or specialized data processing tools. Consequently, we propose the following method for merging two citation datasets:


\begin{enumerate}[leftmargin=11pt]
    \item \textbf{Record Matching:} Here, we propose a four-step procedure: 
    \begin{enumerate}[leftmargin=7pt]
            \item \textit{Design a query:} Let us consider two tuples $\pi_{1i} \in D_1$ and $\pi_{2j}\in D_2$ such that they address the same publication $p_k\in P$. Ideally, their publication data attribute values such as titles, journal titles, author lists, etc. must be the same. The aim of this step is to find a/n (set of) attribute pair(s) such that $(a_{1x}, a_{2y}): a_{1x} \in A_1,  a_{2y} \in A_2$ together or individually address the same publication $p_k\in P$ when their values are "equal" for publication records $p_{1i} \in P_1$ and $p_{2j}\in P_2$. For citation datasets, it is challenging to find an exact match for the non-indexed text based attributes. Hence, while designing the query, we also define an equivalence relation over similarity of the attribute values based on the attributes' domain and context.
    \item \textit{Identify the data slices:} Suppose we have designed a query in  involving the set $Q$ of attribute pair(s). Here, we investigate existence of another (set of) attribute pair(s) $(a_{1u}, a_{2v})$, $ a_{1u} \in A_1, a_{2v} \in A_2$ such that $(a_{1u}, a_{2v}) \notin Q$ and values of attribute pair(s) in $Q$ are equal for $p_{1i} \in P_1$ and $p_{2j}\in P_2$ only if a particular relation exists between $p_{1i}^{a_{1u}}$ and $ p_{2j}^{a_{2v}}$, including but not limited to, inequality or equality. If there are multiple pairs in the data slice, the defined relations together apply to the values of pairwise attributes, i.e., with an (and) operator. Based on the domain of attributes in the data slice, we define the relation accordingly based on the degree of similarity.
    \item \textit{Extract the data:} Dealing with big data, we only extract the required data attributes in this step. This means, we only extract the attribute pairs in the set $Q$ and the attributes required for slicing data. In this step, we preprocess the attributes required for the query so that the records are stored in separate files for each data slice. In this way, we can compare the values of the attribute pairs in $Q$ for the records in each slide. This reduces the number of comparison and facilitates parallel processing of the queries.
    \item \textit{Perform the query:}  We perform the query $Q$ over the data slices to find out the matching records in each slice. Then, we join the results from all slices to obtain the joined set $J$ by query $Q$ where $J\subseteq D_1\bigcap D_2$.
    \item \textit{Update the datasets:} We delete matched records from the datasets, i.e., $D_1:=D_1\setminus J$ and $D_2:=D_2\setminus J$.
    \end{enumerate}
    \item \textbf{Assign MUIDs}: We assign new unique identifiers, called MUIDs, to the merged dataset. This ensures the uniqueness of the records in \( D \).
    \item\textbf{Reference Matching:} First, we construct a reference list for each publication record in the merged dataset using MUIDs. In this step, we utilize data slicing and parallel processing to effectively detect whether a reference exists in the merged dataset. If so, we assign the corresponding MUID to the reference list, otherwise, we omit the reference. Next, we deduplicate the references of the articles that exist in $D_1\bigcap D_2$.
\end{enumerate}
To address the challenges of big data, we incorporate several improvement strategies to enable our method to function efficiently without requiring specialized database management systems or big data frameworks. These strategies include: (1) iterative implementation of steps in record matching, where progressively complex queries are executed, with matching records removed after each iteration to streamline subsequent queries. (2) data slicing, which reduces the number of comparisons by eliminating redundant records. (3) data extraction as needed, where only the required data attributes are extracted at each step based on query definitions and data slices, emphasizing the importance of metadata comprehension. Finally, (4) parallel processing is adopted for data extraction by batch reading/writing and filtering/merging data in slices.
\section{Implementation and Results}\label{Sec:Results}
\begin{figure}[ht]
\centering
\includegraphics[trim={0cm 0cm 0cm 0cm},clip,width=0.8\textwidth]{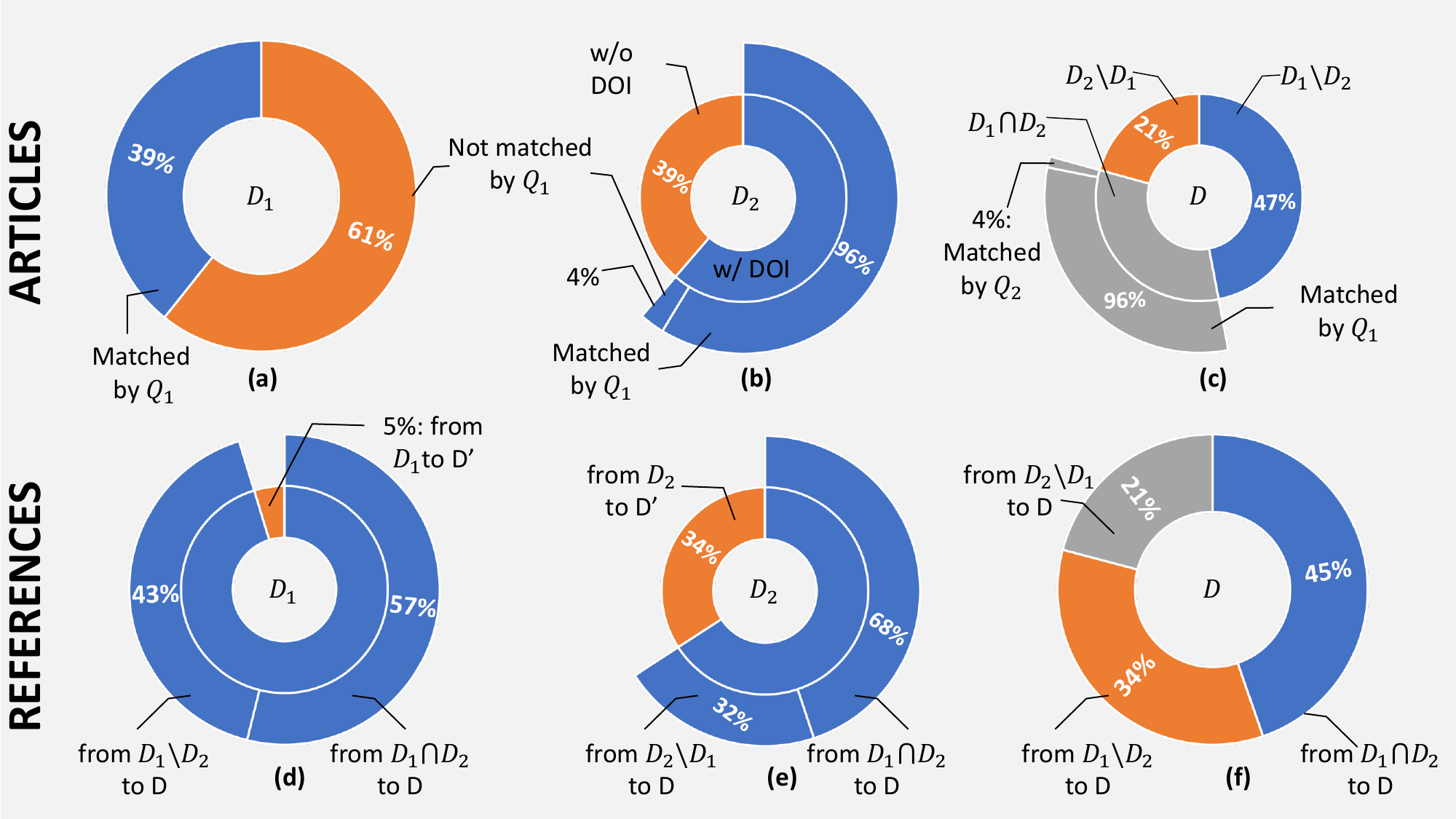}
\caption{Results of merging datasets} \label{Figure_ArtRefA}
\end{figure}

We implemented the proposed method to merge two substantially large research datasets, $D_1$ and $D_2$. $D_1$ is an open database spanning years 1900-2022. It uses DOI as the UID to index article records, but the references are not obliged to include DOI. $D_2$, is a proprietary database spanning years 1981-2020, using its own UID. $D_2$ includes DOI as a data attribute, but not all records in the dataset have a DOI (see Fig. \ref{Figure_ArtRefA} (b)).

We performed record matching by two queries: $Q_1$, $Q_2$. According to $D_1$ and $D_2$'s metadata, publication date-related attributes enable data slicing. First, we aimed to compare the DOIs, being the most reliable attribute for matching the records. We do not expect any records with different publication dates addressing the same article unless there is a quality issue in the input data. We dealt with the exceptional cases by comparing unmatched DOIs after $Q_1$ without slicing. To assess $Q_1$'s accuracy, we compared titles of records with matching DOIs by regular expressions and Levenshtein distance. A 95\% similarity threshold resulted in a 94.19\% match. $Q_2$ matched records by the publication title and journal ISSNs with the same data slicing attributes. Then, we eliminated inconsistent records, i.e., those with titles like \textit{Review} or \textit{Correction}, constituting 0.16\% of the records of the merged dataset, $D$. After record matching, we assigned MUIDs and matched references using parallel processing and data slicing based on MUIDs and UIDs of $D_1$ and $D_2$. Comparing matched records' titles per year, we achieved a 98.86\% match of MUIDs, showing our method's accuracy.



Fig. \ref{Figure_ArtRefA} (a) and (b) show the ratio of matched articles after $Q_1$. Although $D_1$ has DOIs for all records, only 61\% of records in $D_2$ have DOIs. Among those, 4\% of records could not be matched to DOIs in $D_1$. (c) demonstrates the distribution of the articles in $D$ to $D_1$ and $D_2$ and the ratio of records matched by $Q_1$ and $Q_2$. (c) and (d) present the ratio of references in $D_1$ and $D_2$ given to the articles not included by $D$, denoted by $D^{'}$. (f) shows the distribution of edges in the citation graph to the input datasets and their intersection. 

\begin{figure}
		\captionsetup[subfigure]{justification=centering}
		\centering
		\begin{subfigure}{0.49\textwidth}
			\includegraphics[width=1\textwidth]{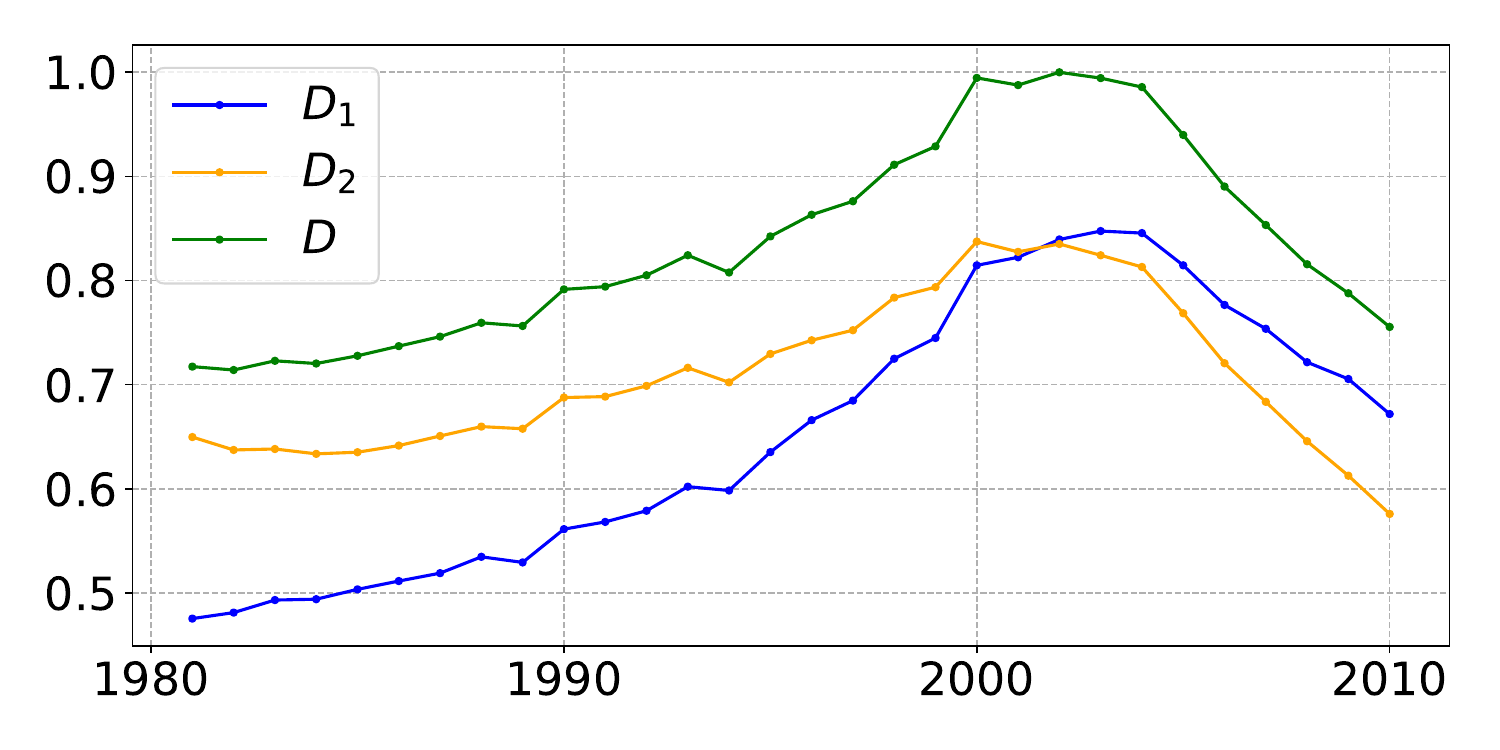} 
			\caption{Average number of citations}
			\label{fig:subim1}
		\end{subfigure}
		\begin{subfigure}{0.49\textwidth}
			\includegraphics[width=1\textwidth]{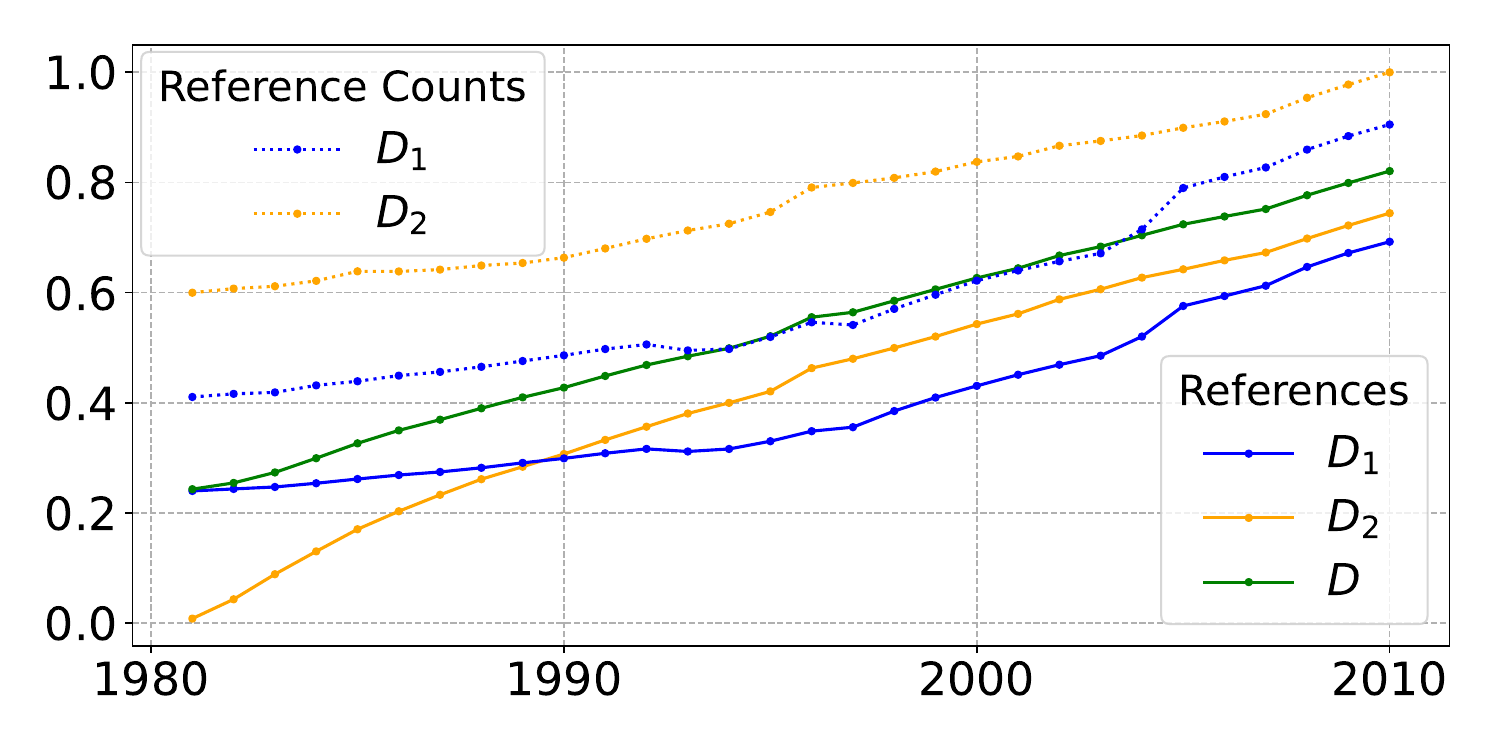}
			\caption{Average number of references}
			\label{fig:subim2}
		\end{subfigure}
		\caption{Comparison of citation and reference counts of matching records}
		\label{fig:AverageRefCit}
	\end{figure}

We also checked the improvement in matching records' references and citations. Fig. \ref{fig:AverageRefCit} (a) and (b) show the comparison of average number of citations and references, respectively, over matched records having at least one reference or citation. Both these plots confirm that merging datasets improves the citation graph. (b) also shows the reference counts provided by the datasets as a separate data attribute. Though the number of references include references other than scientific articles, like datasets, personal communication, etc., (b) shows further improvement opportunity of the citation graph by merging other datasets.
\section{Conclusion}
We present a method for merging large citation datasets to construct comprehensive citation graphs. This method, applicable without specialized database tools or big data frameworks, successfully merged two datasets, leading to a graph with over 119 million nodes and 1.45 billion edges. Merging datasets increased the number of nodes (edges) by 26\% and 89\% (26\% and 52\%) compared to $D_1$ and $D_2$ alone, respectively. Moreover, our results highlight the potential for further improvement through merging additional datasets.
\begin{credits}
\subsubsection{\ackname} This work has been co-funded by the European Union (European Regional Development Fund EFRE, fund number: STIIV-001) and has been partly conducted in the Research Campus MODAL funded by the German Federal Ministry of Education and Research (BMBF) (Fund Nos. 05M14ZAM, 05M20ZBM).
\subsubsection{\discintname}
The authors have no competing interests to declare that are relevant to the content of this article. 
\end{credits}
%
%
%
%

\end{document}